\documentclass[twocolumn,amsfonts,amsmath,amssymb,final,longbibliography,aps,prd,footinbib,nofootinbib,10pt]{revtex4-1}
\usepackage[dvipsnames]{xcolor}
\usepackage{dcolumn}
\usepackage{lipsum}
\usepackage[colorlinks=true,allcolors=Blue]{hyperref}
\usepackage{upgreek,graphicx,manfnt,pifont,colonequals,bbm,stmaryrd}
\usepackage[shortlabels]{enumitem}
\usepackage{mathtools}

\usepackage{amstext}
\usepackage{dsfont} 

\def\be{\begin{eqnarray}}
\def\ee{\end{eqnarray}}
\def\bea{\begin{eqnarray}}
\def\eea{\end{eqnarray}}

\newcommand{\Tr}{\mbox{Tr}}

\newcommand{\beal}{\begin{equation}
\begin{aligned}}
\newcommand{\eeal}{\end{aligned}
\end{equation}}
\newcommand{\bem}{\begin{multline}}
\newcommand{\eem}{\end{multline}}

\def\U{\text{U}}
\def\SU{\text{SU}}
\def\USp{\text{USp}}
\def\SO{\text{SO}}
\def\D{\text{D}}

\begin{document}

\title{Star shaped quivers in four dimensions}

\medskip

\author{Hee-Cheol Kim${}^{a,b}$ and Shlomo S. Razamat${}^c$}
 \email{heecheol@postech.ac.kr, razamat@physics.technion.ac.il }

 \medskip

  \medskip

\affiliation{${}^{a}$ Department of Physics, POSTECH, Pohang 790-784, Korea}
\affiliation{${}^{b}$ Asia Pacific Center for Theoretical Physics, Postech, Pohang 37673, Korea}
\affiliation{${}^{c}$ Department of Physics, Technion, Haifa 32000, Israel}

 \date{\today}


\begin{abstract}
\noindent We discuss a $4d$ Lagrangian descriptions, across dimensions IR dual, of compactifications of the $6d$ $(\D,\D)$ minimal conformal matter theory on a sphere with arbitrary number of punctures and a particular value of flux as a gauge theory with a simple gauge group. The Lagrangian has the form of a  ``star shaped quiver'' with the rank of the central node depending on the $6d$ theory and the number and type of punctures. Using this Lagrangian one can construct across dimensions duals for arbitrary compactifications (any, genus, any number and type of $\USp$ punctures, and any flux)  of the $(\D,\D)$ minimal conformal matter
gauging only symmetries which are manifest in the UV. 

\end{abstract}

\maketitle

\noindent{\bf{Introduction: }}
The notion of infra-red (IR) dualities across dimensions describes the situation when two ultra-violet (UV) QFTs in different dimensions flow in the IR to the same QFT. See \cite{Razamat:2022gpm} for a recent review. As usual, one typically does not rigorously derive such dualities, as these involve strong coupling physics, but rather
constructs a network of self-consistent conjectures. For example, one can consider compactifications of  strongly coupled CFTs on  Riemann surfaces on one hand and dual QFTs defined explicitly in two dimensions less. The latter theories then are naturally labeled by the geometric data of the compactification. This geometric labeling often leads to a geometric understanding of  properties, such as in-dimension IR/conformal dualities and  emergence of symmetry of the lower dimensional theories.

In this note we generalize some of the known across dimensions dualities \cite{Kim:2017toz,Kim:2018bpg,Razamat:2020bix,Nazzal:2021tiu} to a large network of such dualities. In particular, we construct $4d$ Lagrangian duals of compactifications of $(\D_{N+3},\D_{N+3})$ minimal conformal matter theories on spheres with punctures. Such theories are expected to have conformal manifolds with S-duality groups acting on them exchanging the various punctures. Our construction has manifest symmetry under exchanging the punctures and thus describes directly the S-duality invariant locus of the conformal manifold. Moreover we obtain duals to sphere compactifications with more than two maximal punctures with puncture symmetries manifest in the UV. Using these one can construct across dimensions duals to compactifications on surfaces of arbitrary genera with no need 
to rely on gauging of emergent symmetries, contrary to other constructions \cite{Razamat:2020bix,Razamat:2019ukg}. Moreover, for general genus,  values  of flux, and numbers of punctures the $4d$ Lagrangian theory has the structure of a ``{\it star shaped quiver theory}'', reminiscent of $3d$ Lagrangians for class ${\cal S}$ compactifications \cite{Benini:2010uu}. Contrary to the latter case the rank of the central gauge node depends not just on the data of the $6d$ theory but also on the topology of the compactification surface.
\begin{figure}[htbp]
	\centering
  	\includegraphics[scale=0.3]{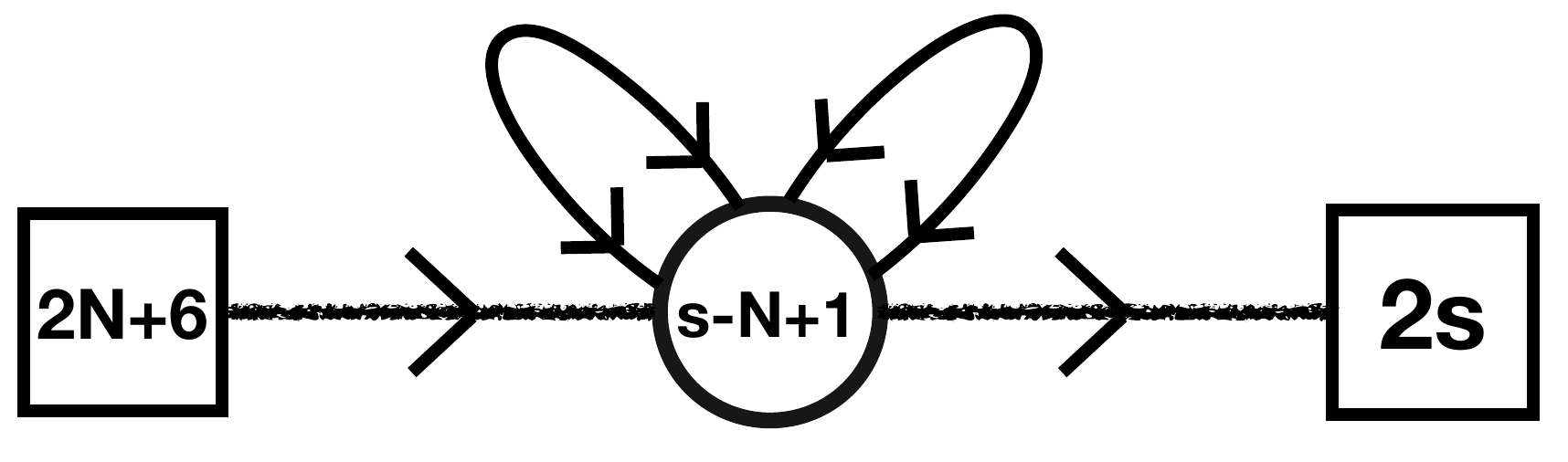}
    \caption{The gauge theory across dimensions dual to compactification on a sphere with $s$ minimal punctures of the $(\D_{N+3},\,\D_{N+3})$ minimal conformal matter theory.}
    \label{F:Gauge}
\end{figure}

\ 

\begin{figure}[htbp]
	\centering
  	\includegraphics[scale=0.25]{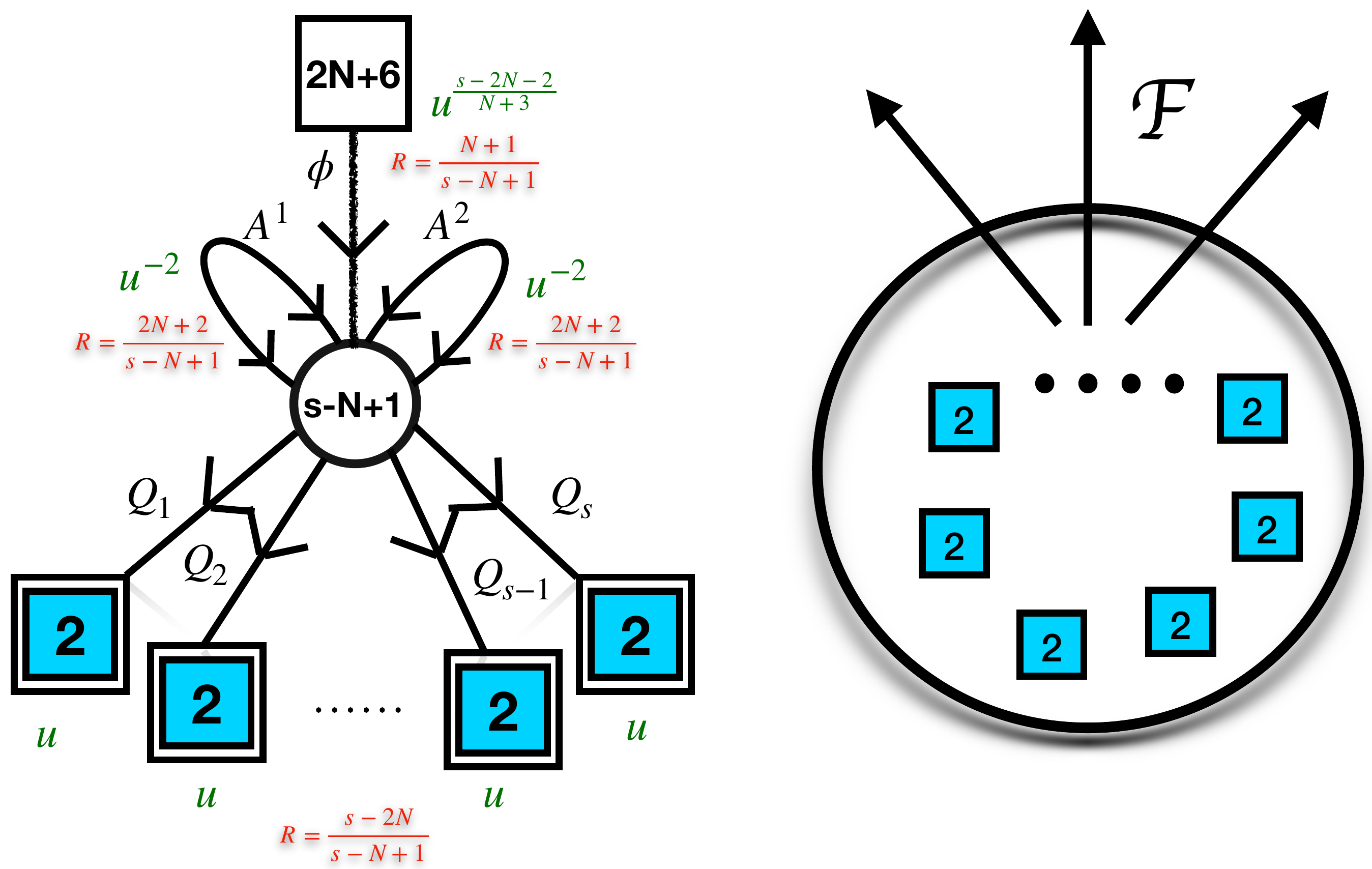}
    \caption{The $2s$ anti-fundamental fields of the theory can be split into $s$ pairs with $\SU(2)$ symmetry. Each pair is associated with a minimal puncture.
    In addition we have $\textcolor{blue}{\U(1)_u\times \SU(2N+6)}$ global symmetry. The values denoted by $\textcolor{red}{R}$ are $6d$ R-charge assignments.}
    \label{F:Estringsphere}
\end{figure}

\begin{figure}[htbp]
	\centering
  	\includegraphics[scale=0.25]{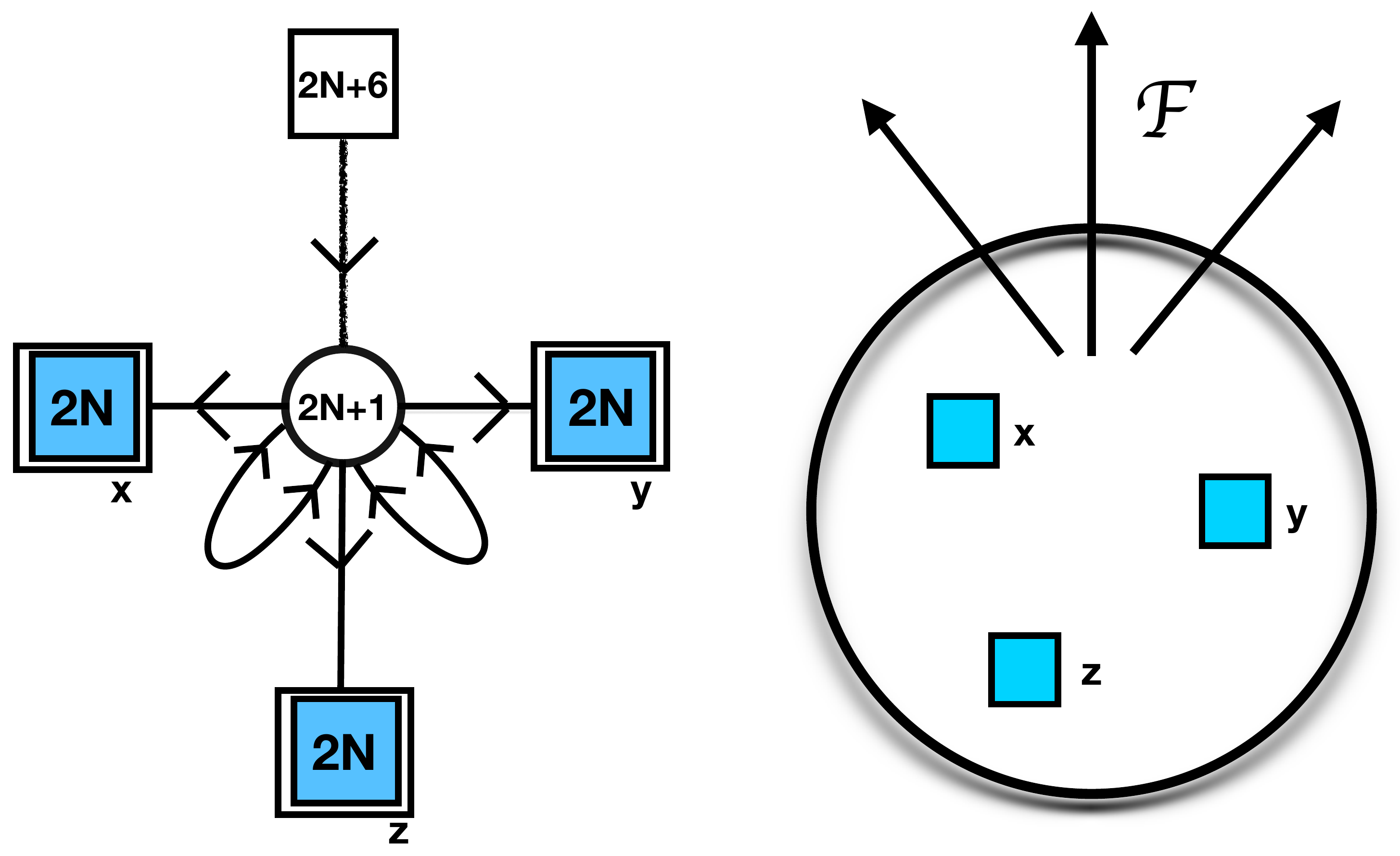}
    \caption{An across dimensions dual of compactification on a three punctured sphere.}
    \label{F:Dtrinion}
\end{figure}

\noindent{\bf{D-type conformal matter on a sphere: }}
We consider a sequence of $6d$ $(1,0)$ conformal theories, the $(\D_{N+3},\D_{N+3})$ minimal conformal matter~\cite{Heckman:2015bfa}, compactified on a Riemann surface.
These $6d$ theories are engineered by studying the low energy dynamics of a single $M5$-brane probing a $\D_{N+3}$ singularity in M-theory. We refer the reader to 
\cite{Ohmori:2014kda,Kim:2018bpg,Razamat:2019ukg} for  discussions of facts about these $6d$ SCFTs we will quote below.
We make the following conjecture which we will next thoroughly explore.  The theory across dimensions dual to the compactification of $(\D_{N+3},\D_{N+3})$ minimal conformal matter on a sphere with $s$ minimal $\SU(2)$ punctures, and certain value of flux to be discussed momentarily, {\it  is the $\SU(s+N-1)$ gauge theory with $2N+6$ fundamental fields $\Phi_i$, $2s$ anti-fundamental fields $Q^a_j$ ($a=1,\cdots , s-N+1$, $j=1,\cdots, 2s$,  two fields in two index anti-symmetric representation $A^{1,2}$, and a superpotential,}
\be\label{symmsup}
W=\sum_{I=1}^2h_I^{ij}\, Q^a_i\,Q^b_j \, A^I_{ab}\,,
\ee with $h_I^{ij}$ being generic coupling constants. See Figures \ref{F:Gauge} and \ref{F:Estringsphere}. The symmetry preserved by this superpotential for general  couplings is $\textcolor{blue}{\U(1)_u\times \SU(2N+6)}\times \SU(2)^s$. The $\textcolor{blue}{\U(1)_u\times \SU(2N+6)}$ part of the symmetry is a subgroup of the global $G_{6d}=\SO(4N+12)$ symmetry of the SCFT in six dimensions.  Each $\SU(2)$ factor will be associated to a minimal puncture. The minimal punctures are defined by a $5d$ limit of the $6d$ SCFT such that the $5d$ effective field theory description is  a $\USp(2N)$ gauge theory, in terms of which we consider boundary conditions at the puncture preserving $\SU(2)\subset \USp(2N)$.  The theory has an  $s-3$ dimensional conformal manifold corresponding to the complex structure moduli of the compactification surface. On this conformal manifold there are  special loci where the symmetry enhances to various proper subgroups of $\USp(2s)$. We expect the theory to have an action of S-duality group exchanging the punctures. We can think of loci with enhanced symmetry as loci of collision of punctures.\footnote{See \cite{Chacaltana:2012ch} for similar effects in class ${\cal S}$ \cite{Gaiotto:2009we,Gaiotto:2009hg}.}
 Whenever $\ell$ punctures collide the symmetry is enhanced as $\SU(2)^\ell \to \USp(2\ell)$. Starting with the maximal puncture, with $\USp(2N)$ symmetry corresponding to the $5d$ gauge group, and partially closing it by Higgs branch flows, one can obtain punctures with symmetry  $\USp(2n)$ for $n\in\{1\cdots N\}$. In particular thus the above theory can describe spheres with any number and any types of $\USp$ punctures.\footnote{
 There are also $\SU(N+1)$ and $\SU(2)^N$ types of maximal punctures \cite{Hayashi:2015fsa,Kim:2018bpg,Kim:2018lfo,Razamat:2019ukg} that we will not discuss here.}
  For example taking $s=3N$ it describes the sphere theory with three maximal punctures. See Figure \ref{F:Dtrinion}.
 Next we discuss  arguments in favor of this conjecture. The special case of $s=2$ and $N=1$ was discussed in \cite{Kim:2017toz},  $s=2N$ and general $N$ in \cite{Kim:2018bpg}, $s=3$ and  $N=1$ in \cite{Razamat:2020bix}, and $s=4$ with $N=1$ in \cite{Nazzal:2021tiu}.

\ 

\noindent{\bf{The conformal manifold: }}
Let us analyze the conformal manifold of the theory in Figure \ref{F:Gauge}. For this we need to determine first the superconformal R-symmetry  as we have an abelian factor of global symmetry, $\U(1)_u$.
We start with the $\SU(s-N+1)$ gauge theory without turning on the superpotential. This gauge theory is non-anomalous and asymptotically free,
\be
\Tr\, R_{free} \SU(s-N+1)^2 = \frac{s+1-3N}3> 0\,,
\ee given that $s\geq 3N$. In particular for a   sphere with three maximal punctures, $s=3N$, we have an asymptotically free theory. 
The theory has two non-anomalous $\U(1)$ symmetries, $\U(1)_u$ we defined above and $\U(1)_v$ which can be chosen such that the fundamentals have charge $\frac1{2N+6}$
while anti-fundamentals have charge $-\frac1{2s}$. One can perform a-maximization \cite{Intriligator:2003jj} and find that the two symmetries mix with the $6d$ R-symmetry (see Figure \ref{F:Estringsphere} for the  the $6d$ R-symmetry assignment).
The superconformal R-charges of all the fields are above $\frac13$ and below $\frac23$ for all the relevant ranges of the parameters $s$ and $N$.
In particular  the superpotential \eqref{symmsup}  is a relevant deformation. These relevant deformations are in ${\bf 2}_A\otimes {\bf A}_{\SU(2s)}$, where  ${\bf A}_{\SU(2s)}$ is the two index antisymmetric irrep of $\SU(2s)$ and ${\bf 2}_A$ is the fundamental irrep of $\SU(2)_A$ symmetry rotating the two fields in the antisymmetric representation.
 We turn on first the superpotential with one antisymmetric field, say $A^1$, and then the other one. 
 At each step one would need to perform a-maximization to determine the superconformal R-symmetry.
The first superpotential breaks the Cartan of the $\SU(2)_A$ and we  have two abelian symmetries which can admix with the R-symmetry. 
The $\SU(2s)$ symmetry is broken to $\USp(2s)$ and as ${\bf Adj}_{\SU(2s)}={\bf A}_{\USp(2s)}+{\bf S}_{\USp(2s)}$ we are left with a relevant operator in ${\bf A}_{\USp(2s)}$, no marginal operators, and symmetry $\U(1)^2\times \USp(2s)\times \SU(2N+6)$.
Turning on next the relevant operator in ${\bf A}_{\USp(2s)}$ we necessarily break $\USp(2s)$ to a subgroup, and generically to $\SU(2)^s$. 
After the dust settles we are left with a conformal manifold of dimension $s-3$ on generic locus of which $\textcolor{blue}{\U(1)_u\times  \SU(2N+6)}\times \SU(2)^s$ symmetry is preserved.
  We can however  preserve a more general subgroup, $\otimes_{i=1}^n \USp(2s_i)$ provided $\sum_{i=1}^ns_i=s$, on sub-loci of the conformal manifold. Insisting on turning on only deformations preserving the above symmetries
  we will end up with \be s-3-\sum_{i=1}^n(s_i-1)=n-3\,,\ee  exactly marginal deformations and marginal deformations breaking this symmetry which are in anti-symmetric irrep for each factor of $\otimes_{i=1}^n \USp(2s_i)$.
The number of exactly marginal deformations preserving the symmetry is what one would expect from having $n$ punctures (with symmetries $\USp(2s_i)$).
In particular note that cases of  $n=1$ or $n=2$ are special as then the dimension of the conformal manifold would have become negative. However, this is not a problem if our theory would be IR free, as happens when $s<3N$. 
In particular note that taking $s=2N$, $n=2$, and $s_1=s_2=N$ the model is an $\SU(N+1)$ gauge theory suggested to correspond to two punctured spheres in \cite{Kim:2018bpg}.
Let us also note that the symmetries of the puncture above can be $\USp(2n)$ with $n>N$.
Let us refer to such punctures as {\it supramaximal} ones.\footnote{It would be interesting to understand these types of punctures better: we suspect these are related to 
obtaining a given $6d$ SCFT as a Higgs branch flow starting from a different one (see {\it e.g.} \cite{Razamat:2019mdt}).} 

\ 

\noindent{\bf{Gluings and flux: }} 
Given the conjecture above first we can construct across dimensions dual to a sphere with three maximal punctures by taking $s=3N$.
The corresponding theory is depicted in Figure \ref{F:Dtrinion}. Using this three punctured sphere theory we can construct across dimensions duals
of compactifications on surfaces of any topology. Note that the sphere theory has a natural set of {\it moment map} operators~\cite{Razamat:2022gpm}, $M = Q\cdot \phi$, which have the following properties: 
they have $6d$ R-charge equal to one; they are in the fundamental irrep of the puncture symmetry; and they are in the fundamental irrep of $\SU(2N+6)$. Each component of the moment map operator $M$ is 
charged under a Cartan of the $6d$ symmetry $G_{6d}=\SO(4N+12)$. 
We can glue surfaces together along  two maximal punctures by gauging the diagonal combination of the symmetries associated to the two punctures and turning on a superpotential.
There is a choice of a superpotential,
\be
W= \sum_{i\in S} M_iM'_i+\sum_{i\notin S}\Phi_i\left(M_i-M_i'\right)\,.
\ee Here $S$ is a subset of the $2N+6$ components of the moment maps, $M$ and $M'$ are the moment maps of the two glued punctures, and $\Phi$s are chiral fields in the fundamental
representation of $\USp(2N)$. When $S$ contains all the moment maps the gluing is called S-gluing and when it is an empty set we call it $\Phi$-gluing. The type of gluing  determines whether the fluxes corresponding
to the $\U(1)$ symmetry of a component  of the moment map are added (in the case of $\Phi$-gluing) or subtracted (in the case of S-gluing).

S-gluing two spheres with $g+1$ maximal punctures together we obtain a genus $g$ Riemann surface with vanishing flux.
 As the following `t Hooft anomaly of the puncture symmetry is,
\be
&&\Tr \U(1)_R\, \USp(2N)^2 =\\&&\qquad  \frac12\left(\frac{s-2N}{s-N+1}-1\right)(s-N+1) = -\frac{N+1}2\,,\nonumber
\ee the gauging is not anomalous. As we glue two copies of the same theory together we also do not have a Witten anomaly.
Since the charges under the $\U(1)$ symmetry of the two glued spheres are opposite, the $\U(1)_u$ symmetry does not mix with the R-symmetry. The conformal anomalies are then simply determined from the 
$6d$ R-symmetries of the various fields,
\be 
a=\frac{3}{16} (g-1) N (16 N+9)\,,\, c=\frac{1}{8} (g-1) N (25 N+18)\,,\nonumber 
\ee which matches perfectly with anomalies of $(\D_{N+3},\,\D_{N+3})$ conformal matter compactified on genus $g$ surface with no flux \cite{Ohmori:2014kda,Kim:2018bpg,Razamat:2019ukg}.
Let us next analyze the dynamics of S-gluing. Taking a sphere with $s$ minimal punctures and performing a-maximization, we obtain the following mixing of $\U(1)_u$ with the R-symmetry,
\be 
&&R_{sc}=R-q_u\frac{N+3}{3 (N-s-1) (4 N-s+8)}\times \nonumber\\
&&\left(\sqrt{2} \sqrt{38 N^2-25 N s+44 N+5 s^2-5 s+8}+6 N-3 s\right)\,.\nonumber
\ee The anomaly $\Tr\, R_{sc}\USp(2N)^2$ is less than $-\frac{N+1}2$ for $s<10N+2$, vanishes for $s=10N+2$, and larger than $-\frac{N+1}2$ for $s>10N+2$. This means that the gauging of $\USp(2N)$ is relevant for $s<10N+2$, marginal for $s=10N+2$, and irrelevant for $s>10N+2$. On the other hand the R-charge of the moment maps is bigger than $1$ for $s<10N+2$, exactly $1$ for $s=10N+2$, and smaller than $1$ for $s>10N+2$. This means that turning on the S-gluing superpotential coupling the moment maps is irrelevant for $s<10N+2$, marginal for $s=10N+2$, and relevant for $s>10N+2$. Thus for $s<10N+2$ we can first perform the gauging, after which the  superpotential will become exactly marginal and then turn it on. 
For $s>10N+2$ we first turn on the superpotential after which the gauging becomes exactly marginal and we perform it. For $s=10N+2$ turning on the superpotential and the gauging together makes them exactly marginal: note that the superpotential and the gauge coupling are charged under the same anomalous symmetry with opposite signs \cite{Green:2010da}. The dynamics of S-gluing always leads to an SCFT. After S-gluing a pair of punctures of the two spheres the gaugings and superpotentials involved in gluing the remaining punctures become exactly marginal.

Finally we would want also to $\Phi$-glue two spheres together to obtain the value of flux one should turn on compactifying on a punctured sphere to obtain the theory in Figure \ref{F:Gauge}.
$\Phi$-gluing two spheres with $g+1$ maximal punctures one obtains genus $g$ surface with the flux being twice the flux of the spheres. Here the charges of the moment maps of the glued punctures are identified
and thus there is a need to perform a-maximization to determine the superconformal R-symmetry.
Doing so one obtains that the anomalies are consistent with the genus $g$ surface having one unit of flux preserving $\U(1)\times \SU(2N+6)$ subgroup of the $\SO(4N+12)$ symmetry of the six dimensional theory.\footnote{
Saying that the flux  breaking $\SO(4N+12)$ to $\U(1)\times \SU(2N+6)$ does not uniquely fix it: there are two choices of $\U(1)$s, corresponding to the choice of a spinorial node in the Dynkin diagram, which do that.
Denoting the flux as a vector of fluxes for the Cartan generators of $\SO(4N+12)$ one obtains $(\pm\frac12,\cdots,\pm\frac12)$ with even or odd number of minus signs depending on the choice of the spinorial node.
One can fix this ambiguity by studying the irreps  appearing in the superconformal index. Doing so one finds that the choice of the spinorial node alternates between odd and even punctures.
See Appendix B for some details.}

We can compute the superconformal index \cite{Kinney:2005ej} both for S-gluing and  $\Phi$-gluing and we find the following result: say for $N=2$ and genus $g$ building the surface from $2g-2$ three punctured spheres,
\begin{widetext}
\be
&&S:\,1+q\,p\left(3g-3+(g-1)(1+{\bf 99}+{\bf 45}u^2+\overline{\bf 45}u^{-2})\right)+\cdots\\
&&\Phi:\,1+q\,p\biggl(3g-3+(g-1)(1+{\bf 99})+(g-1+2g-2)){\bf 45}u^{-2}+(g-1-2g+2)\overline{\bf 45}u^2)\biggr)+\cdots\,,\nonumber
\ee
\end{widetext}
which is consistent with the S-gluing having flux zero and $\Phi$-gluing having flux $1-g$ \cite{babuip}.\footnote{We use the standard notations for the index \cite{Razamat:2022gpm}. The numbers in boldface are $\SU(10)$ irreps.} In particular we observe that in S-gluing the $U(1)_u\times \SU(2N+6)$ symmetry of the Lagrangian enhances to $G_{6d}=\SO(2N+12)$ as expected.
Let us also comment on the dynamics of the $\Phi$-gluing of two spheres. We need to add $2N+6$  fields $\Phi$ in fundamental irrep of $\USp(2N)$ for each puncture and turn on a cubic superpotentials coupling these to moment maps.
As the moment map R-charges are close to one, the superpotential is always relevant and we first turn it on. 
The superpotentials identify all the symmetries of the two glued theories and then the gaugings becomes exactly marginal. 

\ 

\noindent{\bf{RG Flows and Dualities: }}
In addition to gluing surfaces along punctures we can consider closing punctures. The field theoretic procedure corresponding to closing a minimal puncture is as follows, see {\it e.g.} \cite{Razamat:2019ukg}. 
First, we give a VEV (vacuum expectation value) to a component of the moment map operator, $M_i^a=\Phi_i Q^a$ which is a bi-fundamental of $\SU(2N+6)\times \USp(2s)$ where $i$ and $a$ are the indices for the fundamentals of $\SU(2N+6)$ and $\USp(2s)$ respectively.
The VEV breaks the flavor symmetry down to $\USp(2(s-1))$ and also the gauge symmetry to $\SU(s-N+1) \rightarrow \SU(s-N)$. At the end of the RG-flow, we will have the theory of a sphere with $s-1$ minimal punctures  together with extra gauge singlet fields corresponding to Goldstone modes for the broken symmetries which need to be removed by adding certain flip fields. Along the RG-flow, the flavor symmetry $\SU(2N+6)$ is restored as the anti-symmetric fields $A^{1,2}$ decompose into two anti-symmetric fields and two fundamentals for the IR $\SU(s-N)$ gauge symmetry and one combination of these two fundamentals provides an additional fundamental field, {\it i.e.} $\Phi_{2N+6}$, while the other combination of the two fundamentals becomes massive due to the superpotential \eqref{symmsup} with non-zero VEV of the field $Q^s$. Thus, the RG-flow indeed closes a minimal puncture leaving the sphere theory with $s-1$ minimal punctures. See Appendix A for more details.

The  theory across dimensions dual to spheres with punctures should possess conformal dualities exchanging the positions of punctures on the surface.
Note that the field theory we constructed is manifestly invariant under such exchanges of the $\SU(2)$ symmetry factors and thus it flows to the locus on conformal manifold invariant under the duality.
In particular all the supersymmetric partition functions computed for this theory will be manifestly invariant under exchanging the $\SU(2)$ factors. 
We can take a sphere with $s=2g\,N+n$ and $\Phi$-glue pairs of punctures to form a surface of arbitrary genus $g$ and arbitrary number of minimal punctures $n$.
This quiver theory will be ``star-shaped'' and the central node is $\SU((2g-1)N+s-1)$. Moreover, we can change the value of flux by gluing in two punctured spheres. Again we obtain a Lagrangian description manifestly invariant under the dualities 
exchanging punctures which is reminiscent of $3d$ {\it ``star-shaped''} Lagrangians of \cite{Benini:2010uu}.

\ 

\noindent{\bf{{Summary: }}} We have constructed here explicit across-dimensions duals to {\it all} compactifications of a {\it sequence} of $6d$ SCFTs. 
In our construction of duals for general surfaces we do not need to gauge emergent symmetries. Various expected dualities are manifest. It would be interesting
to understand whether similar $4d$ {\it ``star-shaped''} constructions can be obtained for across dimensions duals of compactifciations of other examples of $6d$ SCFTs.

\

\noindent{\bf Acknowledgments}:~
We are grateful to Belal Nazzal, Anton Nedelin, and Gabi Zafrir  for useful discussions.
HK is supported by Samsung Science and Technology Foundation under Project Number SSTF-BA2002-05 and by the National Research Foundation of Korea (NRF) grant funded by the Korea government (MSIT) (No. 2018R1D1A1B07042934).
The research of SSR is supported in part by Israel Science Foundation under grant no. 2289/18, grant no. 2159/22, by a Grant No. I-1515-303./2019 from the GIF, the German-Israeli Foundation for Scientific Research and Development,  by BSF grant no. 2018204. We are grateful to the Aspen Center of Physics for hospitality during initial stages of the project (SSR) and to the Simons Center for Geometry and Physics (HK, SSR). 

\newpage

\appendix

\noindent{\bf Appendix A: RG flow of closing minimal puncture} 
We illustrate the RG-flows of the sphere theory with $s$ punctures that close punctures using superconformal indices \cite{Gaiotto:2012xa,Razamat:2022gpm}. The index of the UV theory is given by,
\begin{widetext}
\be
  \mathcal{I} = && \frac{\left((q;q)_\infty(p;p)_\infty\right)^{s-N}}{(s-N)!}\oint \prod_{\alpha=1}^{s-N}\frac{dt_\alpha}{2\pi i t_\alpha}\frac{\prod_{\alpha<\beta}^{s-N+1}\Gamma_e((pq)^{\frac{N+1}{s-N+1}}u^{-2}t_\alpha t_\beta)^2}{\prod_{\alpha\neq \beta}^{s-N+1}\Gamma_e(t_\alpha/t_\beta)}  \times \nonumber \\
  &&\prod_{\alpha=1}^{s-N+1}\left[\prod_{i=1}^{2N+6}\Gamma_e\left((pq)^{\frac{N+1}{2(s-N+1)}}u^{\frac{s-2N-2}{N+3}}t_\alpha/\mu_i\right) \prod_{a=1}^{s}\Gamma_e\left((pq)^{\frac{s-2N}{2(s-N+1)}}uz_a^{\pm1}/t_\alpha\right)\right] \ , 
\ee
\end{widetext}
where $\mu_i, z_a$ denote the fugacities for the $\SU(2N+6)\times \USp(2s)$ flavor symmetry and $\prod_{i=1}^{2N+6}\mu_i = 1$. The contour is on the unit circle $|t_\alpha|=1$ and we used a shorthand notation $f(z^{\pm1})\equiv f(z)f(z^{-1})$. The definitions of the elliptic gamma function and the q-Pochhammer symbol are as follows,
\be
  \Gamma_e(z) \equiv \prod_{i,j=0}^\infty\frac{1-p^{i+1}q^{j+1}/z}{1-p^iq^jz} \,, \ (z;p)_\infty\equiv \prod_{i=0}^\infty(1-zp^i) \,\nonumber 
\ee

The closing of a minimal puncture can now be accomplished by giving a VEV to a moment map operator, say $M_{2N+6}^s$. 
The index (after performing the $t_\alpha$ integral) exhibits a pole, arising from the contribution of powers of the moment map operator, when the fugacity $\xi=(pq)^{1/2}u^{\frac{s-N+1}{N+3}}z_s/\mu_{2N+6}$ associated with the moment map component is tuned to $\xi\rightarrow1$. The index of the IR theory at the end point of the RG-flow triggered by the VEV can be obtained by taking the residue of this pole and then removing the contributions from Goldstone modes of the broken flavor symmetries.

The pole at $\xi\rightarrow 1$ appears when the contour of the $t_\alpha$ integration is pinched by the following two singularities in the integrand associated with the chiral fields $\Phi_{2N+6}$ and $Q^s$ respectively,
\be\label{eq:poles}
  && (pq)^{\frac{N+1}{2(s-N+1)}}u^{\frac{s-2N-2}{N+3}}t_{\alpha}/\mu_{2N+6} = 1 \ , \nonumber \\
  && (pq)^{\frac{s-2N}{2(s-N+1)}}uz_s/t_{\alpha}=1 \ .
\ee
Then the residue at the pole $\xi= 1$ can be extracted by removing the divergent factor of $\Gamma_e(1)$ from $M_{2N+6}^s$ and tuning the fugacities $t_{s-N+1}$ and $z_s$ as specified in (\ref{eq:poles}), which we compute as,
\begin{widetext}
\be\label{eq:residue}(q;q)_\infty(p;p)_\infty\times {\rm Res}_{\xi\rightarrow 1}\mathcal{I}
  \rightarrow && \frac{((q;q)_\infty(p;p)_\infty)^{s-N-1}}{(s-N-1)!}\oint \prod_{\alpha=1}^{s-N-1}\frac{dt_\alpha}{2\pi i t_\alpha} \frac{\prod_{\alpha<\beta}^{s-N}\Gamma_e((pq)^{\frac{N+1}{s-N+1}}u^{-2}t_\alpha t_\beta)^2}{\prod_{\alpha\neq \beta}^{s-N}\Gamma_e(t_\alpha/t_\beta)} \nonumber\\
  && \times  \prod_{\alpha=1}^{s-N}\left[\prod_{i=1}^{2N+5}\Gamma_e\left((pq)^{\frac{N+1}{2(s-N+1)}}u^{\frac{s-2N-2}{N+3}}t_\alpha/\mu_i\right) \prod_{a=1}^{s-1}\Gamma_e\left((pq)^{\frac{s-2N}{2(s-N+1)}}uz_a^{\pm1}/t_\alpha\right)\right] \nonumber \\
  && \times \prod_{\alpha=1}^{s-N}\Gamma_e((pq)^{\frac{N+1}{2(s-N+1)}}u^{-\frac{s+4}{N+3}}t_\alpha\mu_{2N+6}) \times \mathcal{I}_{\rm extra} \ .
\ee
\end{widetext}
Here, $\mathcal{I}_{\rm extra}$ given by
\be
  \mathcal{I}_{\rm extra}=&&\prod_{i=1}^{2N+6}\Gamma_e(\mu_{2N+6}/\mu_i) \nonumber \\
  &&\times\prod_{a=1}^{s-1}\Gamma_e((pq)^{1/2}u^{\frac{s-N+1}{N+3}}z_a^{\pm1}/\mu_{2N+6}) \ ,
\ee
is the contribution from the Goldstone modes of the broken flavor symmetry $\SU(2N+6)\times \USp(2s)\rightarrow \SU(2N+5)\times \USp(2s-2)$.
We remove this extra factor by adding flipping fields that are coupled to the Goldstone fields through quadratic superpotentials. The remaining elliptic gamma function factors in the last line of (\ref{eq:residue}) provide the contribution of an additional fundamental field with which the $\SU(2N+6)$ flavor symmetry is restored in the IR.
To make this clearer, one can reparametrize the fugacities in the resulting index as 
\be
  && u\rightarrow \mu_{2N+6}^{\frac{1}{s-N+1}}u^{\frac{(s-N)(N+2)}{(N+3)(s-N+1)}}  \ , \nonumber \\
  &&t_\alpha \rightarrow  (pq)^{\frac{N+1}{2(s-N)(s-N+1)}}\mu_{2N+6}^{\frac{1}{s-N+1}}u^{-\frac{s+3}{(N+3)(s-N+1)}}t_\alpha \ , \nonumber \\
  && \mu_{2N+6} \rightarrow \mu_{2N+6}^{-\frac{N+2}{N+3}}u^{\frac{(s-N)(2N+5)}{(N+3)^2}} \ , 
\ee
while maintaining $\prod_{a=1}^{2N+6}\mu_a=1$.
The final index exactly matches the index of the sphere theory with $s-1$ minimal punctures. This outlines the process of closing a minimal puncture at the level of the index.  Note that other types of punctures for larger values of $N>1$ can be considered as collections of minimal punctures. Therefore, any puncture can be closed by repeating the procedure for closing minimal punctures.

\

\noindent{\bf Appendix B: Two flux choices in E-string compactifications}
There are two  options for fluxes that break the $\SO(4N+12)$ flavor symmetry to $\U(1)\times \SU(2N+6)$ in the compactifications of the $(\D,\D)$ minimal conformal matter.
These options correspond to the choices of a spinorial node in the Dynkin diagram of the $\D_{2N+6}$ algebra.
In this appendix, we present an interesting example of these two choices for the E-string theory compactified on a genus 2 Riemann surface. Remarkably, these two choices lead to two distinct 4d theories with identical 't Hooft anomalies but differing superconformal indices.\footnote{See \cite{Distler:2020tub} for similar effects.}

\begin{figure}[htbp]
  \centering\includegraphics[scale=0.24]{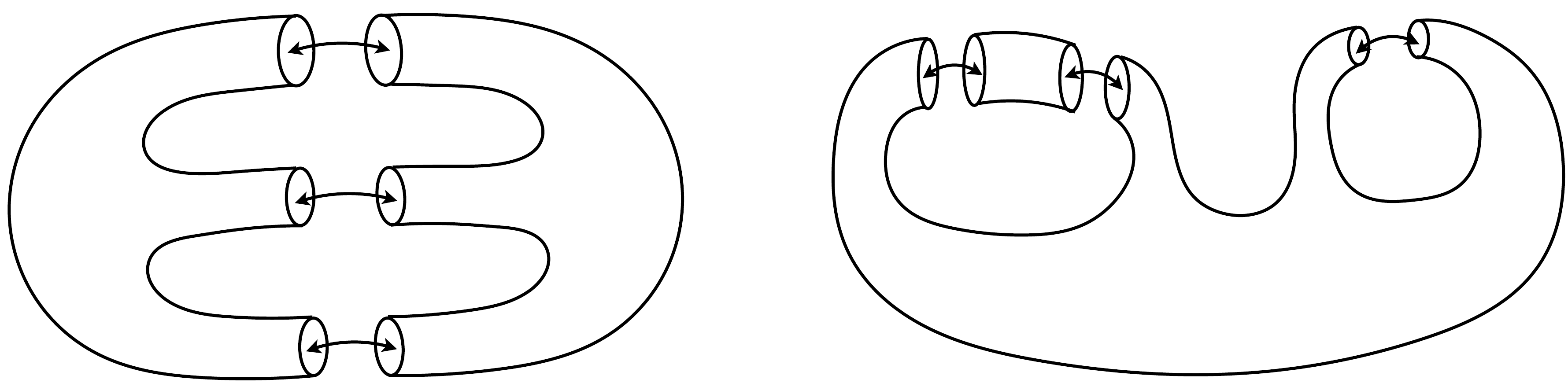}
    \caption{Two constructions for the 6d E-string theory on a genus 2 surface with one unit of flux. The building blocks are the punctured spheres with half-unit flux, and we glue them together using $\Phi$-gluing.
    On the left, the flux is along the co-spinorial node of $\SO(16)$, while on the right, the flux is along the spinorial node. }
    \label{F:genus-2}
\end{figure}

Let us consider two 4d theories arising from the E-string on a genus 2 surface with one unit of such fluxes. The first theory, shown on the left in Figure \ref{F:genus-2}, is constructed by gluing two sphere theories with 3 punctures. Note that the sphere theory with punctures for the E-string theory in Figure \ref{F:Estringsphere} amounts to a half-flux along the spinorial (or co-spinorial) node for $\SO(16)\subset \text{E}_8$ symmetry, depending on whether it has an even (odd) number of punctures.
As the three-punctured sphere theory corresponds to a half-unit flux along the co-spinorial node, this gives rise to the theory obtained from the compactification on a genus 2 surface with a unit flux for the co-spinorial node. On the other hand, the second theory, shown on the right in Figure \ref{F:genus-2}, is obtained by gluing a single sphere theories with 4 punctures to a tube theory (or a two-punctured sphere theory) along one of the punctures, and then gluing two pairs of two punctures. Since all the building blocks have an even number of punctures, the resulting theory corresponds to a genus 2 surface with a unit flux along the spinorial node of $\SO(16)$. We remark that both theories have the same central charges, 
\be
  a = \frac{50+17\sqrt{17}}{16} \ , \quad c=3\frac{19+7\sqrt{17}}{16} \ ,
\ee
and the same 't Hooft anomalies for the global symmetry.

Naively, one might think that the global symmetry of these theories is the $\U(1)\times \SU(8)$ subgroup of the $\text{E}_8$, as it is the maximal subgroup that commutes with the flux. This is indeed the case for the first theory. However, it turns out that the global symmetry of the second theory is enhanced to $\U(1)\times \text{E}_7$. This can be seen, for example, from the 6d $\text{E}_8$ current multiplet and its decomposition into the $\SO(16)$ and $\SU(8)$ irreps. The $\text{E}_8$ current multiplet in ${\bf 248}$ decomposes to ${\bf 120}+{\bf 128}$ representations of $\SO(16)$. The adjoint representation ${\bf 120}$ of $\SO(16)$ further decomposes to $\ {\bf 1}^0 + {\bf 28}^2 + \overline{\bf 28}^{-2} + {\bf 63}^0$ under $\U(1)\times \SU(8)$. There are, however, two distinct decomposisions for the spinorial representation ${\bf 128}$,
\be
  {\bf 128}\ \rightarrow \ {\bf 1}^4 + {\bf 4}^{-4}+ {\bf 28}^{-2} + \overline{\bf 28}^2 + {\bf 70}^0 \ ,
\ee
with the $\U(1)$ associated with the spinorial node of $\SO(16)$ where the superscripts denote the $\U(1)$ charge, and
\be
  {\bf 128}\ \rightarrow \ {\bf 8}^{-3} + \overline{\bf 8}^{3}+ {\bf 56}^{-1} + \overline{\bf 56}^1
\ee
with the $\U(1)$ associated with the co-spinorial node. Under the compactification to 4d with fluxes breaking $\text{E}_8$ to $\U(1)\times \SU(8)$, only the $\U(1)$ neutral multiplets remain as current multiplets. Hence, the above decompositions implies that the global symmetry of the first theory associated with the flux along the co-spinorial node is $\U(1)\times \SU(8)$, while the second theory has a global symmetry that is enhanced to $\U(1)\times \text{E}_7$.
This is because the $U(1)$ neutral multiplets in ${\bf 63}^0 + {\bf 70}^0$  combine to form an adjoint multiplet in ${\bf 133}$ of $\text{E}_7$.

These features of the global symmetry can be verified by computing the superconformal indices. For the first theory, which has the flux along the co-spinorial node, we use the R-symmetry $\U(1)_R^{\rm 6d} + \frac{1}{3}\U(1)_u$, which is close to the superconformal R-symmetry $\frac{-1+\sqrt{17}}{9}-\frac{1}{3}\sim 0.01$, and compute the index as,
\be
  &&\mathcal{I}_{\SU(8)}^{g=2} \;\;\!\!\!\!= 1 + 4u^{-9/4}\chi[\overline{\bf 8}](pq)^{5/8} + 3u^{-3/2}\chi[{\bf 28}](pq)^{3/4} \nonumber \\
  &&\;\;\;\;+2u^{-3/4}\chi[\overline{\bf 56}](pq)^{7/8} + (4+\chi[{\bf 63}])pq + \cdots  .\;\;\
\ee
Here the index is written in terms of the character of the $\SU(8)$ global symmetry. For the second theory with the flux along the spinorial node, we use the R-symmetry $U(1)_R^{\rm 6d} + \frac{2}{7}U(1)_u$, which is also close to the superconformal R-symmetry $\frac{-1+\sqrt{17}}{12}-\frac{2}{7}\sim -0.02$, and compute the index that can now be formed in $\text{E}_7$ characters as
\be
  \mathcal{I}_{\text{E}_7}^{g=2} && = 1 + 5u^{-4} (pq)^{3/7} + 3u^{-2}\chi[{\bf56}](pq)^{5/7} \nonumber \\
  &&+ (4 + \chi[{\bf 133}] )\, p\, q + \cdots \ .
\ee
The coefficients at $(pq)$-order, which takes the form of $3g-3+(g-1) \chi_{\bf adj}$, captures the contributions from the number of marginal operators minus the conserved currents \cite{Beem:2012yn}. The index therefore suggests that the second theory exhibits an enhancement of the global symmetry $\SU(8)\rightarrow \text{E}_7$.

\bibliography{refs}

\end{document}